\documentstyle[12pt]{article}
\def\be{\beta}

\def\ep{\epsilon}

\def\ta{\tau}

\def\om{\omega}

\def\De{\Delta}

\def\fr#1#2{{{#1} \over {#2}}}

\def\pt#1{\phantom{#1}}

\def\frac#1#2{{\textstyle{{#1}\over {#2}}}}

\def\lsim{\mathrel{\rlap{\lower4pt\hbox{\hskip1pt$\sim$}}
    \raise1pt\hbox{$<$}}}
\def\gsim{\mathrel{\rlap{\lower4pt\hbox{\hskip1pt$\sim$}}
    \raise1pt\hbox{$>$}}}
\def\sqr#1#2{{\vcenter{\vbox{\hrule height.#2pt
         \hbox{\vrule width.#2pt height#1pt \kern#1pt
         \vrule width.#2pt}
         \hrule height.#2pt}}}}

\newcommand{\beq}{\begin{equation}}
\newcommand{\eeq}{\end{equation}}
\newcommand{\bea}{\begin{eqnarray}}
\newcommand{\eea}{\end{eqnarray}}
\newcommand{\rf}[1]{(\ref{#1})}

\renewenvironment{thebibliography}[1]
 { \rm
   \begin{list}{\arabic{enumi}.}
    {\usecounter{enumi} \setlength{\parsep}{0pt}
     \setlength{\itemsep}{3pt} \settowidth{\labelwidth}{#1.}
     \sloppy
    }}{\end{list}}

\begin{document}
\titlepage

\begin{flushright}
{IUHET 280\\}
{YCTP-P5-94\\}
{hep-ph/9409404\\}
{June 1994\\}
\end{flushright}
\vglue 1cm

\begin{center}
{{\bf NUCLEAR NULL TESTS FOR SPACELIKE NEUTRINOS
\\}
\vglue 1.0cm
{Alan Chodos$^a$ and V. Alan Kosteleck\'y$^b$ \\}
\bigskip
{\it $^a$Physics Department\\}
\medskip
{\it Yale University\\}
\medskip
{\it New Haven, CT 06511, U.S.A.\\}
\vglue 0.3cm
\bigskip
{\it $^b$Physics Department\\}
\medskip
{\it Indiana University\\}
\medskip
{\it Bloomington, IN 47405, U.S.A.\\}

\vglue 0.8cm
}
\vglue 0.3cm

\end{center}

{\rightskip=3pc\leftskip=3pc\noindent
Recently,
a type of null experiment for spacelike neutrinos
has been proposed.
We examine in detail a class of null tests
involving nuclear beta decay or capture
in atoms and ions.
The most promising candidate systems are identified.

}

\vskip 1truein
\centerline{\it Accepted for publication in Physics Letters B }

\vfill
\newpage

\baselineskip=20pt
{\bf\noindent 1. Introduction}
\vglue 0.4cm

Recently,
we proposed a class of null experiments
designed to test whether the four-momentum
of either the electron neutrino or the muon
neutrino might in fact be spacelike
\cite{ck1,ck2,ko}.
Originally stimulated by theoretical considerations,
this unconventional idea has received further
examination in light of the ongoing trend
towards negative values in the measurement of
squared neutrino masses.
Five recent experiments
\cite{ex1,ex2,ex3,ex4,ex5}
have measured the squared mass of the electron neutrino
to be negative,
with the mean more than two standard deviations from zero.
Measurements of the squared mass of the muon neutrino
over the past decade
\cite{ex6,ex7,ex8,ex9,ex10}
have similarly produced negative values,
with the latest being five standard deviations from zero.
The situation in the muonic case has been altered
very recently by a remeasurement of a crucial pionic x-ray intensity,
which allows a new solution for the pion mass
that yields a muon-neutrino squared mass compatible with zero
\cite{ex11}.
An independent confirmation of this possibility
would be very interesting.
In any event,
the issue is ultimately experimental.
With this in mind,
we present in this letter an analysis of a class
of possible experiments that can provide a \it lower \rm
(negative) bound on the squared mass of the electron neutrino.

Although the focus in this paper is on the electron neutrino,
the basic idea is most easily illustrated for the muon neutrino,
which we summarize first.
More details about the theoretical motivation
and a variety of null tests are presented in
\cite{ck1,ck2,ko}.

Consider the process
\beq
\mu\rightarrow\pi + \nu_\mu
\quad .
\label{a}
\eeq
When the muon is at rest,
this reaction is forbidden by energy
conservation.\footnote{We
disregard here the possibility of negative-energy neutrinos,
which leads to a variety of disagreements with observation.
}
If the neutrino has spacelike momentum,
however,
it is possible to boost the muon to a frame
in which the process is kinematically allowed.
The threshold for the reaction to
occur is given by the muon energy
\bea
E_{\rm th}& = &\fr 1{2\vert m_\nu\vert}
\left[
(m_\pi^2 - m_\mu^2)^2 + 2(m_\pi^2 + m_\mu^2)m_\nu^2 + m_\nu^4
\right]^{1/2}
\nonumber\\
& \simeq & \fr{4.2\times 10^3}{\vert m_\nu/{\rm MeV}\vert}
\quad {\rm MeV}
\quad ,
\label{b}
\eea
where $m_\nu$ is defined to satisfy
$m^2_\nu =p^2_\nu -E^2_\nu$.
A crude estimate of the rate $\om$ of the reaction
\rf{a} can be obtained under a number of theoretical assumptions
\cite{ck2}.
It exhibits a peak at
$E_{\rm max} = \sqrt{3}E_{\rm th}$,
varying with $m_\nu$ and $\bar Q = m_\pi -m_\mu$ as
\beq
\om {(E_{\rm max})}\propto \fr{m_\nu^3}{\bar Q}
\quad ,
\label{d}
\eeq
which makes the rate small.
Nevertheless,
it can be combined with experimental data on muon beams
to place a \it lower \rm (negative) bound on the squared mass of
the muon neutrino
\cite{ko}.

Similar considerations
also apply to processes involving the electron neutrino.
For example,
certain $\be$ decays forbidden kinematically
when the decaying nucleus is at rest
could be allowed if the nucleus were boosted.
More stringent limits have been placed
on the mass of the electron neutrino,
so the corresponding threshold energy
$E_{\rm th}$ might be expected to be higher,
and the corresponding rate $\om (E_{\rm max})$ smaller.
However,
as $m_\mu \rightarrow m_\pi$ for fixed $m_\nu$,
Eqs.\ \rf{b} and \rf{d}
reveal that the threshold energy
becomes independent of $m_\nu$
while the small factor $m_\nu^3$ in the numerator of the rate
can be overcome by the shrinking value of $\bar Q$.
In other words,
if the barrier to the decay at rest is decreased,
then the threshold energy is reduced
while the decay rate is enhanced.

These considerations suggest it is worthwhile to pursue
a search for nuclear $\be$ processes forbidden kinematically
but for which the magnitude $\bar Q$ of
the associated negative $Q$ value is very small.
This paper presents the results of such an investigation
and thereby identifies the
most promising candidates for future experiments.

To ensure a comprehensive search,
it is necessary to consider not only the usual
beta decay involving electrons and antineutrinos,
but also processes like positron emission and electron capture.
Moreover,
since a nuclear null experiment is likely to
involve the acceleration of the nucleus in question,
some or all of the electrons may need to be
stripped from the neutral atom.
This procedure can have a profound
effect on the $Q$ value for the reaction.
For example,
consider the beta decay of
the neutral atom $^{163}_{\pt{1}66}$Dy,
via the process
${^{163}_{\pt{1}66}\rm{Dy}}\rightarrow
{^{163}_{\pt{1}67}\rm{Ho}}+\bar\nu_e$.
This reaction,
which converts one neutral atom to another,
has a small negative $Q$ value,
$Q\simeq -2.6$ keV.
However,
the $Q$ value for beta decay
of the fully stripped dysprosium nucleus
into a holmium ion with one bound electron
is \it positive, \rm
$Q\simeq +50$ keV.
Indeed,
this reaction has recently been experimentally observed
\cite{ju}.
The difference in behavior between
the neutral atom and the fully stripped nucleus
suggests an examination of
intermediate, partially stripped ions
is needed to identify
the most promising candidate reactions
for nuclear null tests.

\vglue 0.6cm
{\bf\noindent 2. Analysis and Results}
\vglue 0.4cm

To investigate these questions systematically,
we begin by introducing a more complete notation.
We use the symbol $^A_Z{\rm{N}}(k)$
to denote a nucleus of element $\rm{N}$
with baryon number $A$, charge $Z$,
and with $k$ electrons bound to it.
The lifetime of the corresponding neutral atom is denoted
by $\ta_{\rm N}$.
The ionicity $n=Z-k$ is
defined as the net positive charge on the ion.
In what follows,
we restrict our attention to the case
$0\le k \le Z$,
i.e.,
we disregard the possibility of negative ions.

Our purpose requires a study of the energetics
of any given process.
For simplicity,
the symbol $^A_Z{\rm{N}}(k)$
is also used to denote the rest energy of a given ion.
The absolute value of the binding energy
of $k$ electrons to a nucleus of charge $Z$
is denoted by $B(k,Z)$.
We then immediately obtain
\bea
^A_Z{\rm{N}}(k)
&=&^A_Z{\rm{N}}(0) + km_e - B(k,Z)
\nonumber\\
&=&^A_Z{\rm{N}}(Z) - nm_e + B(Z,Z) - B(k,Z)
\quad .
\label{f}
\eea
These two formulae relate
$^A_Z{\rm{N}}(k)$
either to the mass $^A_Z{\rm{N}}(0)$ of the bare nucleus
or to the rest energy $^A_Z{\rm{N}}(Z)$ of the neutral atom.

We begin the analysis with a consideration of
processes involving two-body final states.
There are two types of such processes to discuss.
First, we examine reactions of the type
\beq
{^A_Z\rm{N}}_1(k) \rightarrow
{^{\pt{Z+}A}_{Z+1}{\rm{N}}_2(k+1)} + \bar\nu_e
\quad {\rm (type~2a)}
\quad ,
\label{g}
\eeq
in which the nucleus $N_1$ undergoes beta decay
and the emitted electron is captured by the
ensuing ion $N_2$.
Using Eq.\ \rf{f},
the $Q$ values $Q_{2a}(k)$ for this class of processes
can be expressed as
\bea
Q_{2a}(k) & = &
{^A_Z{\rm{N}}_1(Z)}
- {^{\pt{Z+}A}_{Z+1}{\rm{N}}_2(Z)}
+ B(Z,Z) - B(k,Z)
\nonumber\\
& & \qquad - B(Z+1,Z+1) + B(k+1,Z+1)
\nonumber\\
& = & {^A_Z\De} - \De B_Z + \De B_k
\quad .
\label{h}
\eea
In the second form of this equation,
the so-called beta-decay energy ${^A_Z\De}$
is given by
\beq
{^A_Z\De}={^A_Z{\rm{N}}_1(Z)}-{^{\pt{Z+}A}_{Z+1}{\rm{N}}_2(Z+1)}
\quad .
\label{ha}
\eeq
The remaining factors are defined by
\beq
\De B_k = B(k+1,Z+1) - B(k,Z)
\quad ,
\label{hb}
\eeq
which is positive for all $k$.
The quantity
$\De B_Z\equiv \De B_{k=Z}$
is the difference in total electronic binding energy between
the two neutral atoms.

\begin{table}
\centering
\begin{tabular}{|c|c|c|c|} \hline
$^A_Z{\rm{N}_1}$ &
$\ta_{{\rm N}_1}$ &
$^{\pt{Z+}A}_{Z+1}{\rm{N}_2}$ &
$Q_{2a}(Z){\rm ~(keV)}$ \\ \hline
$^{163}_{\pt{1}{66}}$ {\rm Dy} &
 --- &
$^{163}_{\pt{1}{67}}$ {\rm Ho} &
 -2.6 \\ \hline
$^{243}_{\pt{1}{95}}$ {\rm Am} &
$7.4\times 10^3$ y&
$^{243}_{\pt{1}{96}}$ {\rm Cm} &
 -7.3 \\ \hline
$^{202}_{\pt{1}{81}}$ {\rm Tl} &
12.2 d&
$^{202}_{\pt{1}{82}}$ {\rm Pb} &
-46.0 \\ \hline
$^{194}_{\pt{1}{79}}$ {\rm Au} &
38.0 h&
$^{194}_{\pt{1}{80}}$ {\rm Hg} &
-50.0 \\ \hline
$^{123}_{\pt{1}{51}}$ {\rm Sb} &
 --- &
$^{123}_{\pt{1}{52}}$ {\rm Te} &
-52.0 \\ \hline
$^{157}_{\pt{1}{64}}$ {\rm Gd} &
 --- &
$^{157}_{\pt{1}{65}}$ {\rm Tb} &
-57.6 \\ \hline
$^{205}_{\pt{1}{81}}$ {\rm Tl} &
 --- &
$^{205}_{\pt{1}{82}}$ {\rm Pb} &
-60.0 \\ \hline
$^{136}_{\pt{1}{54}}$ {\rm Xe} &
 --- &
$^{136}_{\pt{1}{55}}$ {\rm Cs} &
-67.0 \\ \hline
$^{213}_{\pt{1}{84}}$ {\rm Po} &
4 $\mu$s&
$^{213}_{\pt{1}{85}}$ {\rm At} &
-74.0 \\ \hline
$^{244}_{\pt{1}{94}}$ {\rm Pu} &
$8.1\times 10^7$ y&
$^{244}_{\pt{1}{95}}$ {\rm Am} &
-76.0 \\ \hline
\end{tabular}
\bigskip
\noindent
\caption{Smallest negative
$Q_{2a}(Z)$ values for two-body reactions of type 2a.
}
\end{table}

Let us first consider type-$2a$
reactions with $k=Z$,
i.e., those involving neutral atoms.
For this case,
$\De B_Z = \De B_k$
so we find $Q_{2a}(Z) = {^A_Z\De}$.
Values of ${^A_Z\De}$
can be obtained from a table of atomic masses.
Table 1 shows the most favorable cases
extracted from the compilation in
ref.\ \cite{wb}.
Since some of the candidate atoms are unstable,
we also list their lifetimes $\ta_{{\rm N}_1}$
taken from
ref.\ \cite{nwc}.

The values of $Q_{2a}(k)$ for other nonzero $k$,
i.e.,
the partially stripped cases,
are not readily found.
The problem is that,
to our knowledge,
there is no compendium of data
permitting a determination of
values of $\De B_k$ for nonzero $k$.
Resolving this issue requires an alternative approach,
to which we return below.

Instead,
consider the case $k=0$.
This corresponds to the beta decay of a fully stripped
nucleus into an ion with one bound electron.
The $Q$ value for a process of this type is given
by $Q_{2a}(0) = {^A_Z\De}-\De B_Z + B(1,Z+1)$,
since by definition $B(0,Z)\equiv 0$.
The quantity $\De B_Z$ can be calculated from
tables of total atomic energies
\cite{de},
while
the quantity $B(1,Z+1)$ is provided
in ref.\ \cite{js}.
We have determined the candidate
processes with the most favorable $Q_{2a}(0)$ values
from among the approximately 1500 nuclides
for which data are available.
These processes are listed in Table 2.

\begin{table}
\centering
\begin{tabular}{|c|c|c|c|c|c|c|} \hline
$^A_Z{\rm{N}_1}$ &
$\ta_{{\rm N}_1}$ &
$^{\pt{Z+}A}_{Z+1}{\rm{N}_2}$ &
$^A_Z\De$ &
$\De B_Z$ &
$B(1,Z+1)$ &
$Q_{2a}(0)$ \\ \hline
$^{157}_{\pt{1}{64}}${\rm Gd} &
--- &
$^{157}_{\pt{1}{65}}${\rm Tb} &
-57.6 & 11.9 & 61.0 &- 8.5  \\ \hline
$^{235}_{\pt{1}{92}}${\rm U}{\pt{1}}&
$7.0\times 10^8$ y&
$^{235}_{\pt{1}{93}}${\rm Np} &
-123.0 & 21.6 & 135.2 &- 9.4 \\ \hline
$^{123}_{\pt{1}{51}}${\rm Sb}&
--- &
$^{123}_{\pt{1}{52}}${\rm Te} &
-52.0 & 8.5 & 38.2 &-22.3 \\ \hline
$^{178}_{\pt{1}{73}}${\rm Ta}&
2.4 h&
$^{178}_{\pt{1}{74}}${\rm W{\pt{1}}} &
-89.0 & 14.7 & 80.8 &-22.9 \\ \hline
$^{238}_{\pt{1}{92}}${\rm U}\pt{1} &
$4.5\times 10^9$ y&
$^{238}_{\pt{1}{93}}${\rm Np} &
-145.6 & 21.6 & 135.2 &-32.0 \\ \hline
$^{136}_{\pt{1}{54}}${\rm Xe}&
--- &
$^{136}_{\pt{1}{55}}${\rm Cs} &
-67.0 &  9.2 & 42.9 & -33.3 \\ \hline
$^{176}_{\pt{1}{70}}${\rm Yb}&
--- &
$^{176}_{\pt{1}{71}}${\rm Lu} &
-109.6 & 13.7 & 73.8 & -49.6 \\ \hline
$^{179}_{\pt{1}{72}}${\rm Hf}&
--- &
$^{179}_{\pt{1}{73}}${\rm Ta} &
-115.0 & 14.3 & 78.3 & -51.0 \\ \hline
$^{160}_{\pt{1}{64}}${\rm Gd}&
--- &
$^{160}_{\pt{1}{65}}${\rm Tb} &
-102.3 & 11.9 & 61.1 & -53.2 \\ \hline
$^{ 82}_{34}${\rm Se}&
--- &
$^{ 82}_{35}${\rm Br} &
-88.0 &  4.8  & 16.9 &-75.9 \\ \hline
\end{tabular}
\bigskip
\caption{
Smallest negative $Q_{2a}(0)$ values in keV
for two-body reactions of type 2a.
}
\end{table}

Next, we consider the second type of process
with a two-body final state,
namely,
electron capture (`inverse beta decay').
Typically,
this proceeds from the K shell,
but L- and higher-shell captures can also occur.
The latter are suppressed by wavefunction overlap
but are energetically favored.
Unless an experiment clearly determines the capture type,
the energetics of a null test require the assumption
that the outermost electron is captured.
We therefore consider reactions of the form
\beq
{^{\pt{Z+}A}_{Z+1}{\rm{N}}_2(k+1)}
\rightarrow {^A_Z{\rm{N}}_1(k)} + \nu_e
\quad {\rm (type~2b)}
\quad .
\label{i}
\eeq
We denote the corresponding
$Q$ values by $Q_{2b}(k+1)$,
and using Eqs.\ \rf{f} and \rf{h}
we find
\beq
Q_{2b}(k+1)=-Q_{2a}(k)
\quad .
\label{j}
\eeq

Just as for the case of type-$2a$ processes,
available data permit a numerical study of
reactions with neutral atoms, $k=Z$,
and with fully stripped nuclei, $k=0$.
Since the right-hand side of Eq.\ \rf{j}
has a negative sign,
it suffices to repeat the previous procedures
but keeping now processes with small \it positive \rm $^A_Z\De$
listed in ref.\ \cite{wb}.
The most favorable cases
for $Q_{2b}(Z+1)$ and $Q_{2b}(1)$
are displayed in Tables 3 and 4,
respectively.

\begin{table}
\centering
\begin{tabular}{|c|c|c|c|} \hline
$^{\pt{Z+}A}_{Z+1}{\rm{N}_2}$ &
$\ta_{{\rm N}_2}$ &
$^A_Z{\rm{N}_1}$ &
$Q_{2b}(Z+1){\rm ~(keV)}$ \\ \hline
${^{187}_{\pt{2}76}}$ {\rm Os} &
--- &
${^{187}_{\pt{2}75}}$ {\rm Re} &
 -2.6 \\ \hline
${^{2}_{3}}$ {\rm He} &
--- &
${^{1}_{3}}$ {\rm H} &
 -18.6 \\ \hline
${^{241}_{\pt{2}95}}$ {\rm Am} &
433 y&
${^{241}_{\pt{2}94}}$ {\rm Pu} &
 -20.8 \\ \hline
${^{222}_{\pt{2}87}}$ {\rm Fr} &
14.2 m&
${^{222}_{\pt{2}86}}$ {\rm Rn} &
 -32.0 \\ \hline
${^{148}_{\pt{2}64}}$ {\rm Gd} &
75 y&
${^{148}_{\pt{2}63}}$ {\rm Eu} &
 -33.0 \\ \hline
${^{107}_{\pt{2}47}}$ {\rm Ag} &
--- &
${^{107}_{\pt{2}46}}$ {\rm Pd} &
 -33.1 \\ \hline
${^{250}_{\pt{2}97}}$ {\rm Bk} &
3.2 h&
${^{250}_{\pt{2}96}}$ {\rm Cm} &
 -37.0 \\ \hline
${^{106}_{\pt{2}45}}$ {\rm Rh} &
29.8 s&
${^{106}_{\pt{2}44}}$ {\rm Ru} &
 -39.4 \\ \hline
\end{tabular}
\bigskip
\noindent
\caption{Smallest negative $Q_{2b}(Z+1)$ values
for two-body reactions of type 2b.
}
\end{table}

\begin{table}
\centering
\begin{tabular}{|c|c|c|c|c|c|c|} \hline
$^{\pt{Z+}A}_{Z+1}{\rm{N}_2}$ &
$\ta_{{\rm N}_2}$ &
$^A_Z{\rm{N}_1}$ & $^A_Z\De $ &
$\De B_Z$ &
$B(1,Z+1)$ &
$Q_{2b}(1)$ \\ \hline
$^{215}_{\pt{1}{86}}${\rm Rn} &
2.3 $\mu$s&
$^{215}_{\pt{1}{85}}${\rm At}&
-82.0 &18.8 & 112.8 &-12.1  \\ \hline
$^{213}_{\pt{1}{85}}${\rm At} &
0.11 $\mu$s&
$^{213}_{\pt{1}{84}}${\rm Po}&
-74.0 &18.4 & 109.9 & -17.5 \\ \hline
$^3_2${\rm He} &
--- &
$^3_1${\rm H} &
18.6 & 0.0 & 0.0 & -18.6 \\ \hline
$^{205}_{\pt{1}{82}}${\rm Pb} &
$1.5\times 10^7$ y&
$^{205}_{\pt{1}{81}}${\rm Tl}&
-60.0 &17.4 & 101.3 & -24.0 \\ \hline
$^{194}_{\pt{1}{80}}${\rm Hg} &
520 y&
$^{194}_{\pt{1}{79}}${\rm Au}&
-50.0 &16.7 & 95.9 & -29.2 \\ \hline
$^{202}_{\pt{1}{82}}${\rm Pb} &
$5.3\times 10^4$ y&
$^{202}_{\pt{1}{81}}${\rm Tl}&
-46.0 &17.4 & 101.3 & -38.0 \\ \hline
$^{246}_{\pt{1}{98}}${\rm Cf} &
35.7 h&
$^{246}_{\pt{1}{97}}${\rm Bk}&
-80.0 &23.8 & 153.1 & -49.4 \\ \hline
$^{163}_{\pt{1}{67}}${\rm Ho} &
$4.6\times 10^3$ y&
$^{163}_{\pt{1}{66}}${\rm Dy}&
-2.6 &12.5 & 65.1 & -50.1 \\ \hline
\end{tabular}
\bigskip
\noindent
\caption{Smallest negative $Q_{2b}(1)$ values
in keV for two-body reactions of type 2b.
}
\end{table}

We next turn to a consideration of
processes involving three-body final states.
There are again two types of such processes to examine.
We begin with reactions of the
standard beta-decay type
\beq
{^A_Z{\rm{N}}_1(k)}\rightarrow
{^{\pt{Z+}A}_{Z+1}{\rm{N}}_2(k)} + e^- + \bar\nu_e
\quad {\rm (type~3a)}
\quad .
\label{k}
\eeq
The $Q$ values $Q_{3a}(k)$ for processes
of this type can be expressed
using Eq.\ \rf{f} as
\medskip
\bea
Q_{3a}(k) & = & {^A_Z{\rm{N}}_1(Z)}
- {^{\pt{Z+}A}_{Z+1}{\rm{N}}_2(Z+1)}
+ B(Z,Z)- B(k,Z)
\nonumber\\
& & \qquad - B(Z+1,Z+1) + B(k,Z+1)
\nonumber\\
&=& {^A_Z\De} - \De B_Z + \De \widehat B_k
\quad ,
\label{l}
\eea
where $\De \widehat B_k = B(k,Z+1) - B(k,Z)$
is a non-negative quantity.

There are several cases of type-$3a$ processes
accessible to analysis.
For reactions involving an initially neutral atom, $k=Z$,
the quantity
$-\De B_Z + \De \widehat B_k$
reduces to
\beq
-\De B_Z + \De \widehat B_Z \equiv - I(Z) = - B(Z+1,Z+1) + B(Z,Z+1)
\quad ,
\label{la}
\eeq
where $I(Z)$ is just the (positive)
first-ionization energy of the nuclide N$_2$.
The order of magnitude of $I(Z)$ is one to ten eV,
so to an excellent approximation we can write
$Q_{3a}(Z) \approx {^A_Z\De} = Q_{2a}(Z)$.
A listing of the most favorable such processes
has already been given in Table 1.

\begin{table}
\centering
\begin{tabular}{|c|c|c|c|c|c|c|c|} \hline
$^A_Z{\rm{N}_1}$ &
$\ta_{{\rm N}_1}$ &
$^{\pt{Z+}A}_{Z+1}{\rm{N}_2}$ &
$^A_Z\De $ &
$\De B_Z$ &
$\De\widehat B_1 $ &
$Q_{3a}(1)$ &
$Q_{3a}(0)$ \\ \hline
$^{241}_{\pt{1}{94}}${\rm Pu}
&14.4 y
& $^{241}_{\pt{1}{95}}${\rm Am}
&20.8 &22.4 &3.5 &$>0$ &-1.6  \\ \hline
$^{187}_{\pt{1}{75}}${\rm Re}
& $4.4\times 10^{10}$ y
& $^{187}_{\pt{1}{76}}${\rm Os}
&2.6 &15.3 &2.5 &-10.2 &-12.7  \\ \hline
$^{163}_{\pt{1}{66}}${\rm Dy}
&---
& $^{163}_{\pt{1}{67}}${\rm Ho}
&-2.6 &12.5 &2.0 &-13.1 &-15.1  \\ \hline
$^{243}_{\pt{1}{95}}${\rm Am}
& $7.4\times 10^3$ y
& $^{243}_{\pt{1}{96}}${\rm Cm}
&-7.3 &22.9 &3.6 &-26.5 &-30.1  \\ \hline
$^{202}_{\pt{1}{81}}${\rm Tl}
&12.2 d
& $^{202}_{\pt{1}{82}}${\rm Pb}
&-46.0 &17.4 &2.7 &-60.7 &-63.4  \\ \hline
\end{tabular}
\bigskip
\noindent
\caption{Binding energies and smallest $Q_{3a}(1)$
and $Q_{3a}(0)$ values in keV for three-body reactions of type 3a.
}
\end{table}

As mentioned above for the two-body processes of type $2a$,
we know of no tabulation of data that permit
a determination of $Q_{3a}(k)$ for arbitrary nonzero $k$.
However,
the situation is more favorable
than for type-$2a$ processes because
Eq.\ \rf{l} involves $\De \widehat B_k$ rather than $\De B_k$.
The data compiled in ref.\ \cite{js}
therefore permit an analysis of the case where $k=1$.
The situation for the fully stripped nucleus with $k=0$,
which has $\De \widehat B_k = 0$,
can also be analyzed.
Among the nuclides for which information is available,
we have found only a few candidates of interest
for either $k=0$ or $k=1$.
They are listed in Table 5.
Whenever $Q$ is negative,
the candidate with $k=1$ is more favorable
than the corresponding one for $k=0$
because $\De \widehat B_1$ is positive.

The second class of reaction involving a three-body final
state is positron emission,
with the generic form
\beq
{^{\pt{Z+}A}_{Z+1}{\rm{N}}_2(k)}
\rightarrow {^A_Z{\rm{N}}_1(k)} + e^+ + \nu_e
\quad {\rm (type~3b)}
\quad .
\label{m}
\eeq
We find the associated $Q$ value $Q_{3b}(k)$ is
related to $Q_{3a}(k)$ as given in Eq.\ \rf{l}
by
\beq
Q_{3b}(k)=-Q_{3a}(k)-2m_e
\quad .
\label{o}
\eeq

\begin{table}
\centering
\begin{tabular}{|c|c|c|c|c|} \hline
$^{\pt{Z+}A}_{Z+1}{\rm{N}_2}$ &
$\ta_{{\rm N}_2}$ &
$^A_Z{\rm{N}_1}$ &
$^A_Z\De {\rm ~(keV)}$ &
$Q_{3b}(Z){\rm ~(keV)}$ \\ \hline
$^{185}_{\pt{1}{76}}${\rm Os} &
93.6 d&
$^{185}_{\pt{1}{75}}${\rm Re} &
-1015.0& -7.0 \\ \hline
$^{67}_{31}${\rm Ga} &
3.3 d&
$^{67}_{30}${\rm Zn} &
-1001.1& -20.9 \\ \hline
$^{122}_{\pt{1}{54}}${\rm Xe} &
20.1 h&
$^{122}_{\pt{1}{53}}${\rm I} &
-1000.0& -22.0 \\ \hline
$^{191}_{\pt{1}{78}}${\rm Pt} &
2.9 d&
$^{191}_{\pt{1}{77}}${\rm Ir} &
-1000.0& -22.0 \\ \hline
$^{83}_{37}${\rm Rb} &
86.2 d&
$^{83}_{36}${\rm Kr} &
-998.0& -24.0 \\ \hline
$^{236}_{\pt{1}{93}}${\rm Np} &
$1.2\times 10^5$ y &
$^{236}_{\pt{1}{92}}${\rm U} &
-984.0& -38.0 \\ \hline
$^{238}_{\pt{1}{96}}${\rm Cm} &
2.4 h&
$^{238}_{\pt{1}{95}}${\rm Am} &
-980.0& -42.0 \\ \hline
$^{203}_{\pt{1}{82}}${\rm Pb} &
52.0 h&
$^{203}_{\pt{1}{81}}${\rm Tl} &
-974.0& -48.0 \\ \hline
\end{tabular}
\bigskip
\noindent
\caption{Smallest negative $Q_{3b}(Z)$ values
for three-body reactions of type 3b.
}
\end{table}

As for type-$3a$ processes,
the data available make possible an investigation
of positron emission
for the cases with $k=Z$, $k=1$, and $k=0$.
The presence of the factor of $-2m_e$
on the right-hand side
of Eq.\ \rf{o} means that reactions of
most interest have $^A_Z\De$ values
close to $-2m_e$.
Table 6 lists
favorable cases for $Q_{3b}(Z)$,
while the most interesting cases for $Q_{3b}(1)$
and $Q_{3b}(0)$ are displayed in Table 7.
The table shows that cases with $k=1$ are less
favorable than those with $k=0$,
because the expression for $k=1$ contains in addition the
negative quantity $-\De \widehat B_1$.

\begin{table}
\centering
\begin{tabular}{|c|c|c|c|c|c|c|c|} \hline
$^{\pt{Z+}A}_{Z+1}{\rm{N}}_2$ &
$\ta_{{\rm N}_2}$ &
$^A_Z{\rm{N}}_1$ &
$^A_Z\De$ &
$\De B_Z$ &
$\De \widehat B_1$ &
$Q_{3b}(1)$ &
$Q_{3b}(0)$ \\ \hline
$^{191}_{\pt{1}{78}}${\rm Pt}&
2.9 d&
$^{191}_{\pt{1}{77}}${\rm Ir}&
-1000.0 & 16.0 & 2.6&-8.6& -6.0 \\ \hline
$^{122}_{\pt{1}{54}}${\rm Xe}&
20.1 h&
$^{122}_{\pt{1}{53}}${\rm I}&
-1000.0 & 9.0 & 1.6&-14.6&-13.0 \\ \hline
$^{236}_{\pt{1}{93}}${\rm Np}&
$1.2\times 10^5$ y&
$^{236}_{\pt{1}{92}}${\rm U} &
-984.0 & 21.6 & 3.4&-19.8&-16.4 \\ \hline
$^{67}_{31}${\rm Ga} &
3.3 d&
$^{67}_{30}${\rm Zn}&
-1001.1 & 4.0 & 0.8&-17.7&-16.9 \\ \hline
$^{83}_{37}${\rm Rb}&
86.2 d&
$^{83}_{36}${\rm Kr}&
-998.0  & 5.2 & 1.1&-19.9&-18.8 \\ \hline
$^{238}_{\pt{1}{96}}${\rm Cm}&
2.4 h&
$^{238}_{\pt{1}{95}}${\rm Am}&
-980.0&22.9 & 3.5&-22.6&-19.1 \\ \hline
$^{203}_{\pt{1}{82}}${\rm Pb}&
52.0 h&
$^{203}_{\pt{1}{81}}${\rm Tl}&
-974.0&17.4 & 2.7&-33.4&-30.7 \\ \hline
$^{220}_{\pt{1}{90}}${\rm Th}&
9.7 $\mu$s&
$^{220}_{\pt{1}{89}}${\rm Ac}&
-916.0&20.3 & 3.2&-88.9&-85.7 \\ \hline
$^{76}_{33}${\rm As}&
26.3 h &
$^{76}_{32}${\rm Ge}&
-923.0  & 4.4 & 0.9&-95.6& -94.7\\ \hline
$^{250}_{100}${\rm Fm}&
30 m&
$^{250}_{\pt{1}{99}}${\rm Es}&
-900.0&24.7& 4.0&-101.2&-97.3\\ \hline
\end{tabular}
\bigskip
\noindent
\caption{Smallest negative $Q_{3b}(1)$
and $Q_{3b}(0)$ values in keV for three-body reactions of type 3b.
}
\end{table}

To summarize the above analysis,
we have examined two- and three-body final states
for a variety of nuclear processes
in neutral atoms and fully or almost fully stripped nuclei.
Kinematically,
the most promising candidate
for two-body reactions is the dysprosium-holmium
case with $Q_{2a}(Z)\simeq -2.6$,
while that
for three-body reactions is the plutonium-americium
case with $Q_{3a}(0)\simeq -1.6$.

Let us next return to the issue of partially stripped
ions with $k$ electrons.
A complete analysis of all possible cases
would require a numerical computation of the values
of $B(k,Z)$ for all values of $k$ and $Z$,
a task beyond the scope of this paper.
Instead,
we present here an analysis for a specific case,
the dysprosium-holmium pair.
Intermediate ionicities are of particular interest
for this example not only because
$^A_Z\De$ is small but also because the quantities
$Q_{2a}(k)$ and $Q_{2b}(k+1)$
change sign as one passes from the
neutral dysprosium atom to the fully stripped ion.

\begin{table}
\centering
\begin{tabular}{|c|c|c|c|c|c|} \hline
$n$ &
$B(67-n,67)$ &
$B(66-n,66)$ &
$\De B_{66-n}$ &
$\De \widehat B_{66-n}$ &
$Q_{2a}(66-n)$
\\ \hline
66 &  65.2 & 0.0 & 65.2
& 0.0
& 50.1
\\ \hline
65 & 129.2 &  63.2 & 66.0
& 2.0
& 50.9
\\ \hline
64 & 145.0 & 125.1 & 19.9
& 4.1
& 4.8
\\ \hline
63 & 160.4 & 140.3 & 20.1
& 4.7
& 5.0
\\ \hline
62 & 175.5 & 155.3 & 20.2
& 5.1
& 5.1
\\ \hline
61 & 190.1 & 169.8 & 20.3
& 5.7
& 5.2
\\ \hline
60 & 203.6 & 184.0 & 19.6
& 6.1
& 4.5
\\ \hline
59 & 216.7 & 197.0 & 19.7
& 6.6
& 4.6
\\ \hline
58 & 229.5 & 209.7 & 19.8
& 7.0
& 4.7
\\ \hline
57 & 242.0 & 222.1 & 19.9
& 7.4
& 4.8
\\ \hline
56 & 247.6 & 234.2 & 13.4
& 7.8
& -1.7
\\ \hline
55 & 253.1 & 239.6 & 13.5
& 8.0
& -1.6
\\ \hline
54 & 258.4 & 244.9 & 13.5
& 8.2
& -1.6
\\ \hline
53 & 263.5 & 250.0 & 13.5
& 8.4
& -1.6
\\ \hline
52 & 268.3 & 254.9 & 13.4
& 8.6
& -1.7
\\ \hline
51 & 273.0 & 259.5 & 13.5
& 8.8
& -1.6
\\ \hline
50 & 277.6 & 264.1 & 13.5
& 8.9
& -1.6
\\ \hline
49 & 282.1 & 268.5 & 13.6
& 9.1
& -1.5
\\ \hline
48 & 286.2 & 272.8 & 13.4
& 9.3
& -1.7
\\ \hline
47 & 290.2 & 276.8 & 13.4
& 9.4
& -1.7
\\ \hline
46 & 294.1 & 280.6 & 13.5
& 9.6
& -1.6
\\ \hline
45 & 297.9 & 284.4 & 13.5
& 9.7
& -1.6
\\ \hline
44 & 301.5 & 288.0 & 13.5
& 9.9
& -1.6
\\ \hline
43 & 305.0 & 291.4 & 13.6
& 10.1
& -1.5
\\ \hline
42 & 308.3 & 294.7 & 13.6
& 10.3
& -1.5
\\ \hline
41 & 311.6 & 298.0 & 13.6
& 10.3
& -1.5
\\ \hline
\end{tabular}
\caption{
Binding energies and $Q_{2a}(k)$ values in keV
for holmium and dysprosium ions
(continued on next page).
}
\end{table}

\setcounter{table}{7}

\begin{table}
\centering
\begin{tabular}{|c|c|c|c|c|c|} \hline
$n$ &
$B(67-n,67)$ &
$B(66-n,66)$ &
$\De B_{66-n}$ &
$\De \widehat B_{66-n}$ &
$Q_{2a}(66-n)$
\\ \hline
40 & 314.7 & 301.1 & 13.6
& 10.5
& -1.5
\\ \hline
39 & 317.7 & 304.0 & 13.7
& 10.7
& -1.4
\\ \hline
38 & 319.5 & 306.9 & 12.6
& 10.8
& -2.5
\\ \hline
37 & 321.2 & 308.6 & 12.6
& 10.9
& -2.5
\\ \hline
36 & 322.8 & 310.3 & 12.5
& 10.9
& -2.6
\\ \hline
35 & 324.4 & 311.8 & 12.6
& 11.0
& -2.5
\\ \hline
34 & 325.8 & 313.3 & 12.5
& 11.1
& -2.6
\\ \hline
33 & 327.2 & 314.6 & 12.6
& 11.2
& -2.5
\\ \hline
32 & 328.5 & 315.9 & 12.6
& 11.3
& -2.5
\\ \hline
31 & 329.8 & 317.2 & 12.6
& 11.3
& -2.5
\\ \hline
30 & 331.0 & 318.5 & 12.5
& 11.3
& -2.6
\\ \hline
29 & 332.0 & 319.5 & 12.5
& 11.5
& -2.6
\\ \hline
28 & 333.0 & 320.5 & 12.5
& 11.5
& -2.6
\\ \hline
27 & 334.1 & 321.5 & 12.6
& 11.5
& -2.5
\\ \hline
26 & 335.0 & 322.4 & 12.6
& 11.7
& -2.5
\\ \hline
25 & 335.9 & 323.3 & 12.6
& 11.7
& -2.5
\\ \hline
24 & 336.7 & 324.1 & 12.6
& 11.8
& -2.5
\\ \hline
23 & 337.5 & 324.9 & 12.6
& 11.8
& -2.5
\\ \hline
22 & 338.3 & 325.7 & 12.6
& 11.8
& -2.5
\\ \hline
21 & 339.0 & 326.4 & 12.6
& 11.9
& -2.5
\\ \hline
20 & 339.5 & 327.0 & 12.5
& 12.0
& -2.6
\\ \hline
19 & 340.0 & 327.5 & 12.5
& 12.0
& -2.6
\\ \hline
18 & 340.4 & 327.9 & 12.5
& 12.1
& -2.6
\\ \hline
17 & 340.8 & 328.3 & 12.5
& 12.1
& -2.6
\\ \hline
16 & 341.2 & 328.7 & 12.5
& 12.1
& -2.6
\\ \hline
 0 & 343.4 & 330.9 & 12.5
& 12.5
& -2.6
\\ \hline
\end{tabular}
\caption{
Binding energies and $Q_{2a}(k)$ values in keV
for holmium and dysprosium ions
(continued from previous page).
}
\end{table}

Several codes are available for computing
binding energies.
For the present case,
we need the values of
$B(k,67)$ and $B(k,66)$
for a broad range of $k$.
Results for these binding
energies obtained\footnote{These results
were computed and communicated to us by Farid Parpia.}
using the code GRASP2
\cite{grasp2}
are shown in Table 8.
{}From them,
we can obtain the values of $\De B_k$ and $\De \widehat B_k$,
also shown in Table 8.
Together with the observation that for the dysprosium-holmium
system ${^A_Z\De} - \De B_0 \simeq -15.1$ keV,
these results suffice to determine
$Q_{2a}(k)\simeq -15.1 {\rm ~keV} + \De B_k$,
$Q_{2b}(k+1)=-Q_{2a}(k)$, and
$Q_{3a}(k)\simeq -15.1 {\rm ~keV} + \De \widehat B_k$.
Note that the values of $Q_{3b}(k)$ are not interesting
in the present context because they are dominated by
the factor of $-2m_e$ appearing in Eq.\ \rf{o}.

The results for
$Q_{2a}(k)$ are displayed in the last column of Table 8.
Those for $Q_{2b}(k+1)$ follow by changing the sign.
The table shows the existence of some kinematically favorable
partially stripped cases.
The most attractive one appears when $k=27$,
for which $Q_{2a}(27) \simeq -1.4$ keV,
the smallest value yet found.
For $k=6$, a minimum in $Q_{2b}(7)$ of $-4.5$ keV is manifest.

The values of $Q_{3a}(k)$ are not explicitly presented
in the table because,
as can be seen from the values of $\De \widehat B_k$,
they are all negative.
They have magnitudes decreasing monotonically with increasing $k$,
down to the minimum of $Q_{3a}(66)\simeq -2.6$ keV already found.
Indeed,
the value $Q_{3a}(k)\simeq -2.6$ keV holds
for $k\ge 48$.
As might be expected from the relatively small size of
the binding energies of the outermost electrons
in a neutral atom,
$Q_{3a}(k)$ changes little for $k$ near $Z$.
This means our earlier results for $k=Z$
have wider applicability than just to the neutral atoms.
In principle,
this could provide a means of accelerating effectively neutral
atoms for the purposes of a null test.

\vglue 0.6cm
{\bf\noindent 3. Discussion}
\vglue 0.4cm

We have seen above that there are a number of candidate
systems fulfilling the kinematic criteria for
a null test with relatively small $Q$ values.
For a more definite experimental proposal,
several facts should be borne in mind.

First,
the atomic mass tables we have used \cite{wb}
were originally published over fifteen years ago.
Although in the intervening period
more results have been obtained,
to our knowledge there has been no analogous
recent systematic compilation,
as is required for the present analysis.
In particular,
this means that some of the $Q$ values
could be significantly less than
those we have recorded above.
For this reason,
we have also listed systems
with $Q$ values that,
although still relatively small,
are considerably larger than the smallest ones found.
Similarly,
we have not attempted to estimate errors in the data presented.

Second,
in distinguishing between two-body and three-body
final states,
we are assuming that the experiment can
separate these two cases.
If only the transmutation
of one atomic species into another is detected,
then for the test to be meaningful
\it both \rm relevant $Q$ values must be negative.

Finally,
in the above analysis we have assumed
an ideal case for which the nuclides in question
are in the ground state and have a long enough
half life for an experiment to be performed.
In particular,
we have been primarily concerned
with $Q$ values rather than decay rates.
Determining the latter is
a much more involved task than for the case
$\mu\rightarrow\pi\nu$ mentioned in the introduction,
and such a calculation must await a more
definitive experimental proposal.

Let us briefly address the issue of the
boost required for a null test with a given $Q$ value.
A reasonable guide to the boost needed
can be obtained by approximating a
generalization of the formula \rf{b} for
the threshold energy of the two-body decay
$\mu\rightarrow \pi\nu$.
Consider the general two-body null-test reaction
$X_1\rightarrow X_2\nu$,
and denote by $M_1$ the mass of the parent body
and $M_2>M_1$ the mass of the daughter.
Let $\bar Q$ represent
the modulus of the $Q$ value for
this reaction,
so that $ M_2 = M_1+\bar Q$.
Then, we find
\beq
E_{\rm th}^2 =\fr 1 {4m^2}(\bar Q^2+m^2)
\left[(2M_1+\bar Q)^2+m^2\right]
\quad ,
\label{q}
\eeq
where $m$ denotes the real and positive mass parameter
for the neutrino.

It is certainly true that $m\ll M_1$.
Moreover,
$\bar Q \ll M_1$ for all reactions of interest.
It is therefore a good approximation to write
\beq
E_{\rm th} \approx \fr {M_1\bar Q}{m}
\left( 1+\fr{m^2}{\bar Q^2}\right)^{1/2}
\quad .
\label{t}
\eeq
If $m\ll \bar Q$,
as is true for all processes considered above,
the last factor in parentheses can be neglected.
In the reactions of interest,
$M_1$ represents the mass of a nucleus of baryon number $A$,
so $M_1\approx A$ in GeV.
We finally obtain
\beq
m\approx \fr {\bar Q}{(\ep_{\rm th}/{\rm GeV})}
\quad ,
\label{v}
\eeq
where $\ep_{\rm th}$ is the threshold energy per nucleon
of the parent nucleus.

The most promising cases
we have identifed so far have $\bar Q$
of about one keV.
Plans at CERN and RHIC call for ion beams with energy
of order 100 A GeV.
Formula \rf{v} shows
that such beams are kinematically sensitive
to spacelike neutrinos with
mass parameter $m$ on the order of 10 eV.
This is competitive with current bounds
obtained by other means.

\vglue 0.6cm
{\bf\noindent Acknowledgments}
\vglue 0.4cm

We thank Farid Parpia for computing
the binding energies presented in Table 8.
We benefitted from communications with
Walter Johnson, Bob Pollock, and Buford Price.
V.A.K. thanks the Aspen Center for Physics
for hospitality while part of this investigation
was performed.
This work was supported
in part by the United States Department of Energy
under grant number DE-FG02-91ER40661.

\vglue 0.6cm
{\bf\noindent References}
\vglue 0.4cm

\def\plb #1 #2 #3 {Phys.\ Lett.\ B #1 (19#2) #3.}
\def\mpl #1 #2 #3 {Mod.\ Phys.\ Lett.\ A #1 (19#2) #3.}
\def\prl #1 #2 #3 {Phys.\ Rev.\ Lett.\ #1 (19#2) #3.}
\def\pr #1 #2 #3 {Phys.\ Rev.\ #1 (19#2) #3.}
\def\prd #1 #2 #3 {Phys.\ Rev.\ D #1 (19#2) #3.}
\def\npb #1 #2 #3 {Nucl.\ Phys.\ B#1 (19#2) #3.}
\def\ptp #1 #2 #3 {Prog.\ Theor.\ Phys.\ #1 (19#2) #3.}
\def\jmp #1 #2 #3 {J.\ Math.\ Phys.\ #1 (19#2) #3.}
\def\nat #1 #2 #3 {Nature #1 (19#2) #3.}
\def\prs #1 #2 #3 {Proc.\ Roy.\ Soc.\ (Lon.) A #1 (19#2) #3.}
\def\ajp #1 #2 #3 {Am.\ J.\ Phys.\ #1 (19#2) #3.}
\def\lnc #1 #2 #3 {Lett.\ Nuov.\ Cim. #1 (19#2) #3.}
\def\nc #1 #2 #3 {Nuov.\ Cim.\ A#1 (19#2) #3.}
\def\jpsj #1 #2 #3 {J.\ Phys.\ Soc.\ Japan #1 (19#2) #3.}
\def\ant #1 #2 #3 {At. Dat. Nucl. Dat. Tables #1 (19#2) #3.}
\def\nim #1 #2 #3 {Nucl.\ Instr.\ Meth.\ B#1 (19#2) #3.}

\end{document}